\documentclass[conference]{IEEEtran}
\IEEEoverridecommandlockouts
\usepackage[ruled,lined]{algorithm2e}
\usepackage{setspace}
\usepackage{float}
\usepackage{cite}
\usepackage{graphicx}
\usepackage{epstopdf}
\usepackage{graphicx,epsf,epic,eepic,psfig,latexcad}
\usepackage{epstopdf}
\usepackage{amsmath,amssymb,amsfonts}
\usepackage{algorithmic}
\usepackage{textcomp}
\usepackage{xcolor}
\newcommand\numeq[1]%
  {\stackrel{\scriptscriptstyle(\mkern-1.5mu#1\mkern-1.5mu)}{=}}

\def\BibTeX{{\rm B\kern-.05em{\sc i\kern-.025em b}\kern-.08em
    T\kern-.1667em\lower.7ex\hbox{E}\kern-.125emX}}
\begin{document}

\title{\vspace{-6mm}Unsourced Random Access with a Massive MIMO Receiver Using Multiple Stages of Orthogonal Pilots%\vspace{-2.5mm}
}
%\vspace{-39mm}
\author{\IEEEauthorblockN{Mohammad Javad Ahmadi and Tolga M. Duman}
\IEEEauthorblockA{\textit{Department of Electrical and Electronics Engineering} \\
Bilkent University, Ankara, Turkey \\
\{ahmadi, duman\}@ee.bilkent.edu.tr}\vspace{-08mm}
\thanks{This research is funded by the Scientific and Technological Research Council of Turkey (TUBITAK) under the grant 119E589.}
}
\vspace{-10mm}

\maketitle
	\begin{abstract}
We study the problem of unsourced random access (URA) over Rayleigh block-fading channels with a receiver equipped with multiple antennas. We employ multiple stages of orthogonal pilots, each of which is randomly picked from a codebook. In the proposed scheme, each user encodes its message using a polar code and appends it to the selected pilot sequences to construct its transmitted signal. Accordingly, the received signal consists of superposition of the users' signals each composed of multiple orthogonal pilot parts and a polar coded part. We use an iterative approach for decoding the transmitted messages along with a suitable successive interference cancellation scheme. Performance of the proposed scheme is illustrated via extensive set of simulation results which show that it significantly outperforms the existing approaches for URA over multiple-input multiple-output fading channels.
	\end{abstract}
\vspace{-1mm}
\section{Introduction}
\vspace{-1mm}
Massive multiple-input multiple-output (MIMO) systems achieve high spectral efficiencies, high energy efficiencies, high data rates, and spatial multiplexing gains by creating a massive number of spatial degrees of freedom (DoF) \cite{marzetta2010noncooperative}. The original applications of massive MIMO have been in broadband communications \cite{marzetta2010noncooperative, bjornson2015massive, bjornson2017massive, hoydis2013massive}; however, more recently, it has also been proposed for Internet-of-Things (IoT) networks in which a very large number of devices sporadically transmit data to a common access point. The so-called \emph{unsourced random access} (URA), which is introduced by Polyanskiy in \cite{polyanskiy2017perspective}, is a paradigm suitable for many applications in IoT networks, where the base station (BS) only cares about the transmitted messages, and the identity of the users is not of concern.\\
\indent In URA, all the active users share the same codebook for their transmission, and the per-user
probability of error (PUPE) is adopted as the performance criterion. Many low-complexity coding schemes are devised for URA over a Gaussian multiple-access channel (GMAC) including T-fold slotted ALOHA (SA)~\cite{ordentlich2017low,facenda2020efficient,vem2019user,glebov2019achievability}, sparse codes ~\cite{amalladinne2018coupled,zheng2020polar,tanc2021massive,han2021sparse,ebert2021stochastic}, and random spreading~\cite{pradhan2020polar,ahmadi2021random,pradhan2021ldpc}. However, GMAC is not a fully realistic channel model for wireless communications. Therefore, in \cite{kowshik2019quasi,kowshik2020energy,kowshik2019energy,kowshik2021fundamental}, the synchronous Rayleigh quasi-static fading MAC is investigated, and the asynchronous set-up is considered in \cite{andreev2019low, amalladinne2019asynchronous}. Recently, several studies have also investigated Rayleigh block-fading channels in a massive MIMO setting. In \cite{fengler2021non}, a covariance-based activity detection (AD) algorithm is used to detect the active messages, while \cite{decurninge2020tensor} employs rank-1 tensors constructed from Grassmannian sub-constellations. Furthermore, a pilot-based scheme is introduced in~\cite{fengler2020pilot} where non-orthogonal pilots are employed for detection and channel estimation, and a polar list decoder is used for decoding messages. \\
\indent The coherence blocklength is defined as the period over which the channel coefficients stay constant. As discussed in \cite{fengler2021non}, depending on the environment, the coherence blocklength of wireless systems in the URA setting may vary in the range of $100\leq T_c\leq 2\times 10^4$, where $T_c$ is measured in terms of the number of transmitted symbols. Although the AD algorithm in \cite{fengler2021non} performs well in fast fading (e.g., when $T_c\leq 320$), it is not implementable with larger blocklengths due to run-time complexity scaling with $T_c^2$. In contrast, the schemes in \cite{decurninge2020tensor,fengler2020pilot} work well only in the large-blocklength regimes (e.g., for $T_c=3200$); that is, in a slow fading environment where large blocklengths can be employed, their decoding performance is better than that of \cite{fengler2021non}.\\
\indent In this paper, we propose a URA scheme over MIMO fading channels, employing pilot transmission for user detection and channel estimation, similar to \cite{liu2018massive1,senel2018grant,fengler2020pilot}. Unlike the previous works that use a single non-orthogonal pilot sequence, in the proposed scheme, each user employs multiple stages of orthogonal pilots selected randomly from a codebook. Since the orthogonality of the pilots in different stages removes the interference, the performance of the pilot detection and channel estimation algorithms in the proposed scheme is improved compared to the decoding performance of the schemes using non-orthogonal pilots. We demonstrate that, while the covariance-based AD algorithm in \cite{fengler2021non} suffers from high computational complexity in large blocklengths, and the algorithms in \cite{fengler2020pilot,decurninge2020tensor} do not work well in the short blocklength regime (hence not suitable for fast fading scenarios), the newly proposed algorithm has a superior performance in both short and large blocklength regimes.\\
\indent The paper is organized as follows. Section II presents the system model for the proposed framework. The encoding and decoding schemes are introduced in Section III. In Section IV, various numerical results are given. Finally, Section V provides our conclusions.\\
\indent The following notation is adopted throughout the paper. We denote the set of imaginary numbers by $\mathbb{C}$. $\left[ \mathbf{T}\right]_{(l,:)}$ and $\left[ \mathbf{T}\right]_{(:,l)}$ are the $l$th row and the $l$th column of $\mathbf{T}$, respectively. $\mathrm{Re} \left(\mathbf{t}\right)$ and $\mathrm{Im} \left(\mathbf{t}\right)$ are used for the real and imaginary parts of $\mathbf{t}$; and, the transpose and Hermitian of matrix $\mathbf{T}$ are denoted by $\mathbf{T}^T$ and $\mathbf{T}^H$, respectively. The notation $|.|$ is used for the cardinality of a set, and $\mathbf{I}_M$ denotes an $M\times M$ identity matrix.
%\vspace{-2mm}
\section{System Model}
%\vspace{-3 mm}
We consider an unsourced random access model over a block-fading wireless channel. The BS is equipped with $M$ receiving antennas connected to $K_T$ potential users, for which $K_a$ of them are active in a given frame. Assuming that the channel coherence time is larger than $L$, we divide the length-$n$ time-frame into $S$ slots of length $L$ ($n = SL$). Each active user randomly selects a single slot to transmit $B$ bits of information. In the absence of synchronization errors, the received signal vector corresponding to the $l$th slot at the $m$th antenna is written as
\begin{align}
\mathbf{y}_{m,l} = \sum_{i\in \mathcal{K}_l}^{ }{h_{m,i}\mathbf{x}\left(\mathbf{w}(i)\right)+{\mathbf{z}_{m,l}}},
\label{eqs1}
\end{align}
where $\mathbf{y}_{m,l}\in \mathbb{C}^{1\times L}$, $\mathcal{K}_l$ denotes the set of active user indices available in the $l$th slot, $\mathbf{x}\left(\mathbf{w}(i)\right)\in \mathbb{C}^{1\times L}$ is the encoded and modulated signal  corresponding to the message bit sequence $\mathbf{w}(i)\in \{0,1\}^B$ of user $i$, $h_{m,i}\sim \mathcal{CN}(0, 1)$ is the Rayleigh channel coefficient between the $i$th user and the $m$th receive antenna, and ${\mathbf{z}_{m,l} }\sim \mathcal{CN}(\mathbf{0}, \mathbf{I}_L)$ is the circularly symmetric complex white Gaussian noise vector. Letting $\mathcal{K}_a$ and $\mathcal{L}_d$ denote the set of active user indices and  the list of decoded messages, respectively, the PUPE of the system is defined in terms of the probability of false-alarm, $p_{fa}$, and the probability of missed-detection, $p_{md}$, as
\begin{align}
P_e = p_{fa}+p_{md},
\end{align}
\noindent where 
\begin{align}
p_{md}=\dfrac{1}{K_a}\sum_{i\in\mathcal{K}_a}^{}{ \mathrm{Pr}(\mathbf{w}(i)\notin \mathcal{L}_d)}, \ \ \ \ 
p_{fa} = \mathbb{E}\left\{\dfrac{n_{fa}}{|\mathcal{L}_d|}\right\},
\end{align}
with $n_{fa}$ being the number of decoded messages that were indeed not sent. The energy-per-bit of the system can be written as ${E_b}/{N_0}=LP/B$, where $P$ denotes the average power of each user per channel use. The objective is to minimize the required energy-per-bit for a target PUPE.
\section{The Proposed Scheme}
\subsection{Encoder}
\label{sec3-a}
\begin{figure}[t!]%[h!]
		\centering
		\includegraphics[width=.9\linewidth]{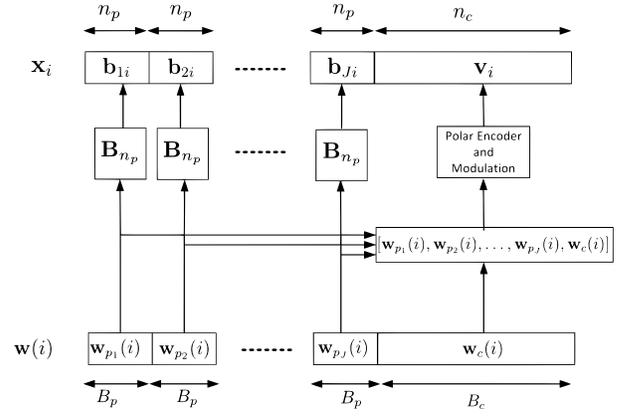}
		\caption{{\small Illustration of the encoding process in the proposed unsourced scheme.}	}	\label{Fig_transmission}
%	\vspace{-6.5 mm}
\end{figure}
 As shown in Fig. \ref{Fig_transmission}, we divide the message of the $i$th user into $J+1$ parts (one data part and $J$ pilot parts) denoted by $\mathbf{w}_{c}(i)$ and $\mathbf{w}_{p_j}(i), j = 1,2,...,J$ with lengths $B_c$ and $B_{p}$, respectively, where $B_c+JB_{p} = B$. The $i$th user obtains its $j$th pilot sequence, $\mathbf{b}_{ji}$, with length $n_{p} = 2^{B_{p}}$ by mapping $\mathbf{w}_{p_j}(i)$ to the orthogonal rows of an $n_{p}\times n_{p}$ Hadamard matrix $\mathbf{B}_{n_{p}}$, which is generated as
	\begin{align}
	\nonumber
	    \mathbf{B}_2  = \begin{bmatrix}1&1\\1&-1\end{bmatrix}, \ \ \ \
	    \mathbf{B}_{2^i} = \mathbf{B}_2 \otimes \mathbf{B}_{2^{i-1}} \ \ \forall \  \ i = 2,3, \hdots, 
    \end{align}
\noindent where $\otimes$ represents the Kronecker product. Since the number of possible pilots in the orthogonal Hadamard codebook is limited, it is likely that the users are in collision in certain pilot segments, that is, they share the same pilots with the other users. However, it is highly unlikely that a given user will experience collision in all the pilot segments. To construct the coded sequence of the $i$th user, we accumulate all the message parts in a row vector as
\begin{align}
 \mathbf{w}(i) = \left[\mathbf{w}_{p_1}(i), \mathbf{w}_{p_2}(i), \hdots, \mathbf{w}_{p_J}(i) ,\mathbf{w}_{c}(i)\right], \label{eq9}
	\end{align}
and pass it to a $\left(2n_c, \  B +r\right)$ polar code, where $r$ is the number of cyclic redundancy check (CRC) bits. Note that contrary to the existing schemes in URA, we feed not only the data bits but also the pilot bit sequences to the encoder. Hence, in the case of successful decoding, all the pilot sequences for the user can be retrieved. The polar codeword is then modulated using quadrature phase shift keying (QPSK), resulting in $\mathbf{v}_i\in \{ \sqrt{{P_{c}}/{2}}(\pm 1\pm j)\}^{1\times n_c }$, where $P_c$ is the average power of the polar coded part. The transmitted signal for the $i$th user also consists of $J$ pilot parts and one coded part as
	\begin{align}
	    \mathbf{x}_i = \left[\sqrt{P_{p}}\mathbf{b}_{1i},\sqrt{P_{p}}\mathbf{b}_{2i},\hdots,\sqrt{P_{p}}\mathbf{b}_{Ji},\mathbf{v}_i\right]\in \mathbb{C}^{1\times L }\label{eq10_old}
	\end{align}
\noindent where $\mathbf{x}_i:=\mathbf{x}\left(\mathbf{w}(i)\right)$, $L = n_c+Jn_{p}$, and $P_p$ denotes the average power of the pilot sequence.\\
\indent The $j$th pilot part and the polar coded part of the received signal in the $l$th slot can be modeled as 
	\begin{align}
\mathbf{Y}_{p_j} &= \sqrt{P_{p}} \mathbf{H}\mathbf{B}_j+\mathbf{Z}_{p_j} \in \mathbb{C}^{M\times n_{p}} , \ j = 1,2,\hdots , J,\label{eq10Pilots}\\
\mathbf{Y}_c &=  \mathbf{H}\mathbf{V}+\mathbf{Z}_c \in \mathbb{C}^{M\times n_c},\label{eq10Polar}
	\end{align}
	where $\mathbf{H}\in \mathbb{C}^{M \times K_l}$ is the channel coefficient matrix with $h_{m,i}$ in its $m$th row and $i$th column, $K_l$ is the number of users in the $l$th slot, $\mathbf{Z}_{p_j}$ and $\mathbf{Z}_c$ consist of independent and identically distributed (i.i.d.) noise samples drawn from $\mathcal{CN}(0,1)$ (i.e., a circularly symmetric complex Gaussian distribution), and the $i$th rows of $\mathbf{B}_j\in \{\pm 1\}^{K_l \times n_{p}}$ and $\mathbf{V}\in \{ \sqrt{{P_{c}}/{2}}(\pm 1\pm j)\}^{K_l \times n_c}$ are $\mathbf{b}_{ji}$ and $\mathbf{v}_i$, respectively. Note that we have removed the slot indices from the above matrices to simplify the notation. 
\subsection{Decoder}	
Decoding in each slot is performed using an iterative process. At each iteration, we decode the transmitted codewords by employing one of the $J$ pilot parts (sequentially) and the coded part of the received signal in \eqref{eq10Pilots} and \eqref{eq10Polar}. Generally, only the non-colliding users can be decoded. Some non-colliding users in the current pilot stage may experience collision in the other pilot transmission parts. Therefore, by successfully decoding and removing them using successive interference cancellation (SIC), the collision density is reduced in the other pilot parts. Repeating the decoding iterations, the effects of such collisions are ameliorated.\\
\indent The decoding process is comprised of five different steps that work in tandem. A pilot detector based on a Neyman-Pearson (NP) test identifies the active pilots in the current pilot part; channel coefficients corresponding to the detected pilots are estimated using a  channel estimator; a maximum-ratio  combining (MRC) estimator is used to produce a soft estimate of the modulated signal; after demodulation, the signal is passed to a polar list decoder; and, the resulting sequences satisfying the CRC are added to the list of successfully decoded signals before being subtracted from the received signal via SIC. The process is repeated until there are no successfully decoded users in $J$ consecutive SIC iterations. In the following, $\mathbf{Y}^\prime_{p_j}$ and $\mathbf{Y}^\prime_{c}$ denote the received signals in \eqref{eq10Pilots} and \eqref{eq10Polar} after removing the list of messages successfully decoded in the current slot up to the current iteration.
\subsubsection{Pilot Detection Based on NP Hypothesis Testing}
At the $j$th pilot part, we can write the following binary hypothesis testing problem:
\begin{align}
\nonumber
     &\mathbf{u}_{ji}|\mathcal{H}_0   \sim \mathcal{CN}\left(\mathbf{0},\mathbf{I}_M\right)
     \\
    & \mathbf{u}_{ji}  |\mathcal{H}_1 \sim \mathcal{CN}\left(\mathbf{0},\sigma^2_1\mathbf{I}_M\right),\label{hypothesis}
\end{align}
where $\sigma_1=\sqrt{1+m_{ij} n_{p}P_{p}}$, $\mathbf{u}_{ji} = \mathbf{Y^\prime}_{p_j}\mathbf{\bar{b}}_{i}^H /\sqrt{n_{p}}$, with $\mathbf{\bar{b}}_{i}=\left[\mathbf{B}_{n_{p}}\right]_{(i,:)}$, $\mathcal{H}_1$ and $\mathcal{H}_0$ are alternative and null hypotheses that show the existence and absence of the pilot $\mathbf{\bar{b}}_{i}$ at the $j$th pilot part, respectively, and $m_{ij}$ is the number of users that pick the pilot $\mathbf{\bar{b}}_{i}$ as their $j$th pilots. Let $\hat{\mathcal{D}}_j$ be the estimate of the set of active rows of $\mathbf{B}_{n_{p}}$ in the $j$th pilot part. Using a $\gamma-$level Neyman-Pearson hypothesis testing (where $\gamma$ is the bound on the false-alarm probability), $\hat{\mathcal{D}}_j$ can be obtained as (see Appendix \ref{appendixC} for details)
\begin{align}
    \hat{\mathcal{D}}_j = \left\{l:\mathbf{u}_{jl}^H\mathbf{u}_{jl} \geq \dfrac{1}{2}\Gamma^{-1}_{2M}(1-\gamma)\right \}\label{eq13},
\end{align}
where $\Gamma_k(.)$ denotes the cumulative distribution function of the chi-squared distribution with $k$ degrees of freedom, and $\Gamma^{-1}_k(.)$ is its inverse. 
%Since the decision rule in \eqref{eq13} does not depend on unknown parameters such as $m$, $n_{p}$ and $P_{p}$, the uniformly most powerful (UMP) test exists for this problem. 
The probability of detection in the absence of collision ($m_{ij}=1$) is obtained in \eqref{pdetect}. Note that a higher probability of detection is obtained in the general case of $m_{ij}> 1 $. It is clear that the probability of detection is controlled by the parameters $\gamma$, $n_{p}$, and $P_{p}$.
\subsubsection{Channel Estimation}
 Let $\mathbf{B}_{\hat{\mathcal{D}}_j}\in \{\pm 1\}^{|\hat{\mathcal{D}}_j|\times n_{p}}$ be a sub-matrix of $\mathbf{B}_{n_{p}}$ consisting of the detected pilots in \eqref{eq13}, and suppose that $\tilde{\mathbf{b}}_{jk}=\left[\mathbf{B}_{\hat{\mathcal{D}}_j}\right]_{(k,:)}$ is the corresponding pilot of the $i$th user. Since the rows of the codebook are orthogonal to each other, the channel coefficient vector of the $i$th user can be estimated as
 \begin{align}
     \hat{\mathbf{h}}_{i}  =\dfrac{1}{ n_{p}\sqrt{P_{p}}}\mathbf{Y^\prime}_{p_j}\tilde{\mathbf{b}}_{jk}^H\label{eq10_Chestimate}.
\end{align} 
\indent Note that if the $i$th user is in a collision (i.e., more than one user selects $\tilde{\mathbf{b}}_{jk}$), Eq.~\ref{eq10_Chestimate} gives an unreliable estimate of the channel coefficient vector. However, this is unimportant since a CRC check is employed after decoding and such errors do not propagate.
\subsubsection{MRC, Demodulation, and Channel Decoding}
Let $\mathbf{h}_{i}$ be the channel coefficient vector of the $i$th user, where $i\in \tilde{\mathcal{S}}_l$ with  $\tilde{\mathcal{S}}_l$ denoting the set of remaining messages in the $l$th slot. Using $\hat{\mathbf{h}}_{i}$ in \eqref{eq10_Chestimate}, the modulated signal of the $i$th user can be estimated employing the MRC technique as
\vspace{-1.0 mm}
\begin{align}
    \hat{\mathbf{v}}_{i} = \hat{\mathbf{h}}_{i}^H\mathbf{Y^\prime}_c. \label{eq17}
\end{align}
 Plugging \eqref{eq10Polar} into \eqref{eq17}, $  \hat{\mathbf{v}}_{i}$ is written as
 \begin{align}
      \hat{\mathbf{v}}_{i}=\hat{\mathbf{h}}_{i}^H\mathbf{h}_i\mathbf{v}_i + \sum_{k\in\tilde{\mathcal{S}}_l, k\neq i}\hat{\mathbf{h}}_{i}^H\mathbf{h}_k\mathbf{v}_k+\hat{\mathbf{h}}_{i}^H\mathbf{Z}_c\label{eq23}.
 \end{align}
 The first term in \eqref{eq23} is the signal term, and the second and third terms are the interference and noise terms, respectively. Since $E\{\mathbf{v}_j^H\mathbf{v}_k\}= P_c \mathbf{I}_{n_c}$, the power of each term can be calculated as
 \begin{align}
    \hat{\sigma}_{si}^2 &\approx P_c  \|\hat{\mathbf{h}}_{i}\|^4,\label{eq25}\\
\hat{\sigma}_{Ii}^2 &= P_c\sum_{k\in \tilde{\mathcal{S}}_l, k\neq i} |\hat{\mathbf{h}}_{i}^H\mathbf{h}_{k}|^2 \approx P_c\sum_{k\in\hat{\mathcal{D}}_j, k\neq i}^{} |\hat{\mathbf{h}}_{i}^H\hat{\mathbf{h}}_{k}|^2 ,\label{eq26}\\
\hat{\sigma}_{ni}^2 & = \|\hat{\mathbf{h}}_{i}\|^2\label{eq27}.
 \end{align}
 \indent Assuming that the interference-plus-noise in \eqref{eq23} is approximately Gaussian-distributed, the following log-likelihood ratio (LLR) is obtained as the input to the polar list decoder
\begin{align}
\mathbf{f}_i = \left[\mathrm{Im}\left(\beta_{1i}\right),\mathrm{Re}\left(\beta_{1i}\right),\hdots,\mathrm{Im}\left(\beta_{n_ci}\right),\mathrm{Re}\left(\beta_{n_ci}\right)\right] \label{eq50LLR},
\end{align}
\begin{align}
\beta_{ti} = \dfrac{ 2\sqrt{\hat{\sigma}_{si}^2}}{\hat{\sigma}_{ni}^2+\hat{\sigma}_{Ii}^2 }\left[\hat{\mathbf{v}}_{i}\right]_{(:,t)}. 
\end{align}
\indent At the $j$th pilot part, the $i$th user is declared as successfully decoded if 1) its decoded message satisfies the CRC check, and 2) by mapping the $j$th pilot part of its decoded message to the Hadamard codebook, $\tilde{\mathbf{b}}_{jk}$ is obtained. Then, all the successfully decoded messages (in the current and previous iterations) are accumulated in the set $\mathcal{S}_l$, where $|\mathcal{S}_l|+|\tilde{\mathcal{S}}_l|=K_l$. 
%\indent At the $l$th slot, all the decoded messages that satisfy the CRC check (in the current and previous iterations) are declared as successfully decoded messages and accumulated in the set $\mathcal{S}_l$, where $|\mathcal{S}_l|+|\tilde{\mathcal{S}}_l|=K_l$. 
\subsubsection{SIC}
\begin{figure*}[t!]%[h!]
		\centering
		\includegraphics[width=.8\linewidth]{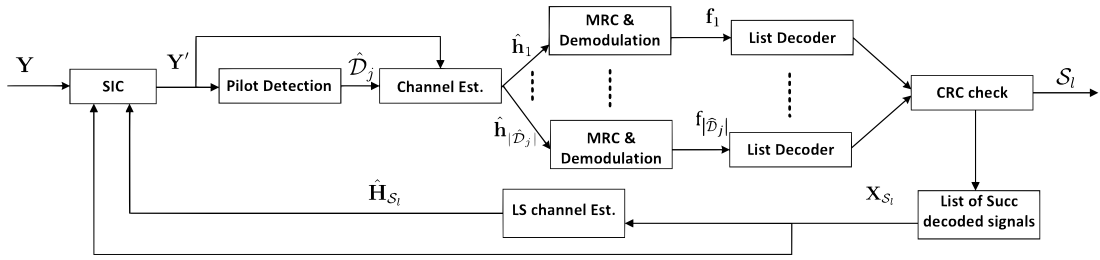}
		\caption{{\small The decoding process at the $j$th pilot part and the $l$th slot.}	}	\label{Fig_decoder}
			\vspace{-5. mm}
\end{figure*}

we can see in \eqref{eq9} that the successfully decoded messages contain bit sequences of pilot parts and the coded part ($\mathbf{w}_{p_j}(i), j = 1,2,...,J$ and $\mathbf{w}_{c}(i)$). Having the bit sequences of successfully decoded messages, we can construct the corresponding transmitted signals using \eqref{eq10_old}. The received signal matrix can be written as
\begin{align}
   \mathbf{Y} =  \mathbf{H}_{\mathcal{S}_l}\mathbf{X}_{\mathcal{S}_l}+\mathbf{H}_{\tilde{\mathcal{S}}_l}\mathbf{X}_{\tilde{\mathcal{S}}_l}+\mathbf{Z}_l, \label{eq21}
\end{align}
  where $\mathbf{Y}$ is obtained by merging received signal matrices of different parts, i.e., $\mathbf{Y} = \left[\mathbf{Y}_{p_1},\hdots,\mathbf{Y}_{p_J}, \mathbf{Y}_c \right]\in \mathbb{C}^{M\times L}$ with $\mathbf{X}_{\mathcal{S}_l}\in \mathbb{C}^{|\mathcal{S}_l|\times L }$ and $\mathbf{X}_{\tilde{\mathcal{S}}_l}\in \mathbb{C}^{|\tilde{\mathcal{S}}_l|\times L }$ being constructed using the signals in the sets $\mathcal{S}_l$ and $\tilde{\mathcal{S}}_l$, and $\mathbf{H}_{\mathcal{S}_l}\in \mathbb{C}^{M \times |\mathcal{S}_l|}$ and $\mathbf{H}_{\tilde{\mathcal{S}}_l}\in \mathbb{C}^{M \times |\tilde{\mathcal{S}_l}|}$ comprise the channel coefficients corresponding to users in the sets $\mathcal{S}_l$ and $\tilde{\mathcal{S}}_l$, respectively. Considering $\mathbf{H}_{\tilde{\mathcal{S}}_l}\mathbf{X}_{\tilde{\mathcal{S}}_l}+\mathbf{Z}_l  $ as an additive noise term, $\mathbf{H}_{\mathcal{S}_l}$ can be estimated by applying the least squares (LS) estimation on \eqref{eq21} as
\begin{align}
    \hat{\mathbf{H}}_{\mathcal{S}_l} =  \mathbf{Y} \mathbf{X}_{\mathcal{S}_l}^H(\mathbf{X}_{\mathcal{S}_l}\mathbf{X}_{\mathcal{S}_l}^H)^{-1}. \label{eq14}
\end{align}
Note that $\mathbf{X}_{\mathcal{S}_l}$ consists of all the successfully decoded signals so far in the $l$th slot, and $\mathbf{Y}$ is the initially received signal matrix, not the output of the latest SIC iteration. The SIC procedure is performed as follows
\begin{align}
    \mathbf{Y^\prime}=\left[\mathbf{Y^\prime}_{p_1}, \mathbf{Y^\prime}_{p_2},\hdots,\mathbf{Y^\prime}_{p_J}, \mathbf{Y^\prime}_c \right] = \mathbf{Y}- \hat{\mathbf{H}}_{\mathcal{S}_l} \mathbf{X}_{\mathcal{S}_l}.\label{eq16}
\end{align}
 Finally, $\mathbf{Y}^\prime$ is fed back to the pilot detection algorithm for the next SIC iteration. The details of decoding stages are shown in Fig. \ref{Fig_decoder} and Algorithm 1.
\begin{algorithm}
	{\small
		\caption{ The proposed decoding scheme.} 
		\For( Different slots){{$l = 0,1, \hdots, S $}}
		{
			$\mathcal{S}_l$    $\ \ =\emptyset$.\\
			
			$\mathrm{flag} \ = 1$.\\
			%	$\mathbf{Y^\prime}_{p_j} = \mathbf{Y}_{p_j}\ , j=1,2, ..., J$. \\
			%	$\mathbf{Y^\prime}_{c}\ =\mathbf{Y}_{c}$.\\
			$t = 0$ (\textit{$t$ shows iteration index}).\\
			\While( {}){$\mathrm{flag}=1$}
			{
				$t \ \ = t+1.$\\
				
				\For(different pilot parts ){{$j = 1, 2,...,J $}}
				{
					\textbf{Pilot detection}:  estimate $\hat{\mathcal{D}}_j$ using \eqref{eq13}.\\
					\For(different detected pilots){{$i \in  \hat{\mathcal{D}}_j$}}
					{
						\textbf{Ch. estimation}:  estimate $\hat{\mathbf{h}}_{i} $ using \eqref{eq10_Chestimate}.\\
						\textbf{Decoding}:  pass $\mathbf{f}_i$ in \eqref{eq50LLR} to list decoder.\\
						$\mathcal{S}_{tj}$: set of successfully decoded users in the current iteration.\\
						$\mathcal{S}_l = 	\mathcal{S}_l \bigcup \mathcal{S}_{tj} $.\\
						\textbf{SIC}:  update $\mathbf{Y^\prime}_{p_j}$ and $\mathbf{Y^\prime}_{c}$ using \eqref{eq16}.
					}
				}
				\If{$\bigcup_{j=1}^J \mathcal{S}_{tj}=\emptyset$}{$\mathrm{flag}=0$.}
			}
		}
	}
\end{algorithm}
\section{Numerical Results}
\begin{figure}[t!]%[h!]
		\centering
		\includegraphics[width=.92\linewidth]{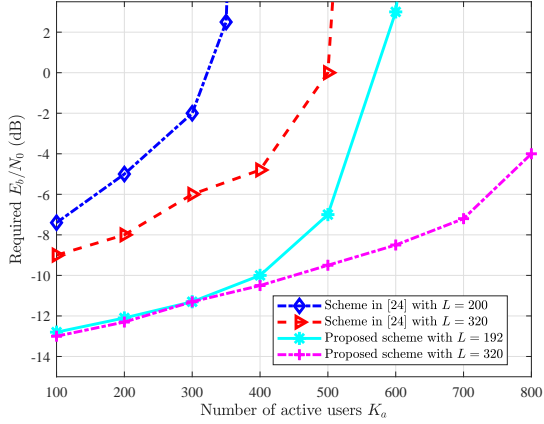}
		\caption{{\small The required $E_b/N_0$ as a function of the number of active users in the proposed scheme and the method in \cite{fengler2021non} for $M=100$ and $L\approx 320, 200$.}	}
		\label{MIMO_short}
	\vspace{-5. mm}
	\end{figure}

\begin{figure}[t!]%[h!]
		\centering
		\includegraphics[width=.92\linewidth]{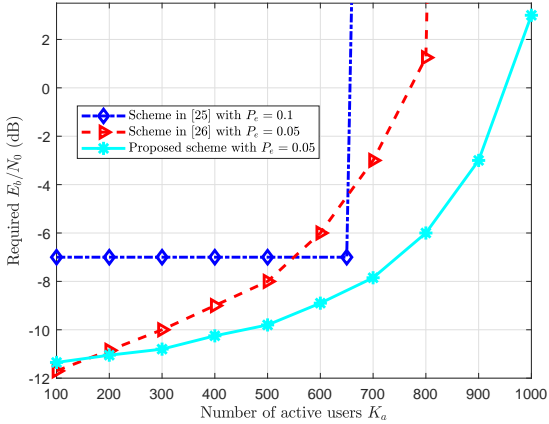}
		\caption{{\small The required $E_b/N_0$ as a function of the number of active users in the proposed scheme and the results in \cite{fengler2020pilot,decurninge2020tensor} for $M=50$.}}	\label{MIMO_long}
	\vspace{-5. mm}
	\end{figure}

\begin{figure}[t!]%[h!]
		\centering
		\includegraphics[width=.92\linewidth]{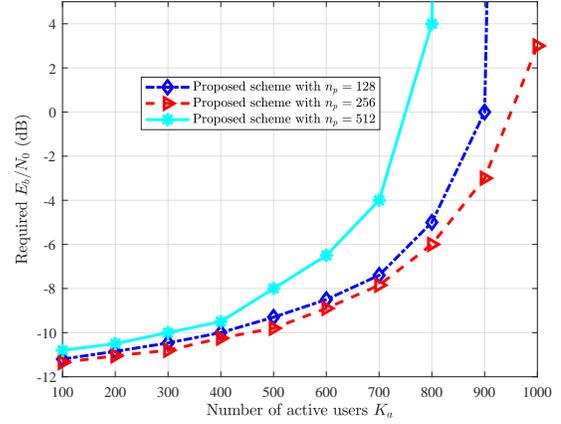}
		\caption{{\small The  required $E_b/N_0$ as  a  function of the number of active users in the proposed scheme for $M=50$, $J = 2$, and different values of $n_p$.}	}	\label{figDiffJ}
	\vspace{-3. mm}
	\end{figure}

\begin{figure}[t!]%[h!]
		\centering
		\includegraphics[width=.92\linewidth]{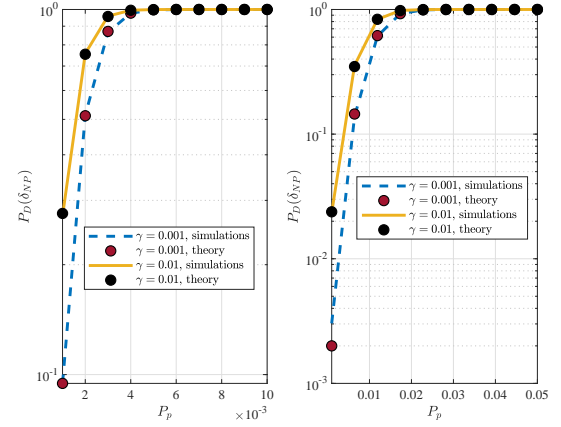}
		\caption{{\small Comparison of the simulation and analytical performance of the pilot detector for $M = 50$ and $n_{p} = 256$ (left), and $M = 100$ and $n_{p} = 32$ (right)}.	}	\label{detectorLong}
			\vspace{-5. mm}
	\end{figure}

In this section, we provide a set of numerical results to assess the performance of the proposed URA scheme. In all the results, we set the list size of the decoder to $64$, $B = 100$, the frame length $n\cong 3200$,  the number of CRC bits $r = 11$, the Neyman-Pearson threshold $\gamma = 0.001$, and $P_e = 0.05$.\\
\indent In Fig. \ref{MIMO_short}, the performance of the proposed scheme is compared with the short blocklength scheme of \cite{fengler2021non} with the number of antennas $M=100$ and slot lengths $L=320$ and $200$. For a fair comparison, we consider two scenarios with $(J,n_p,n_c) = (2,32,256)$ (corresponding to $L=320$) and $P_c/P_p =0.5$, and $(J,n_p,n_c) =  (2,32,128)$ ($L=192$) and $P_c/P_p =1$, respectively. It is illustrated in this figure that the proposed decoder significantly outperforms the approach in \cite{fengler2021non}.\\
\indent To compare the proposed framework with the ones in \cite{decurninge2020tensor} and \cite{fengler2020pilot}, we set $(J,n_p,n_c) = (2,256,512)$, $M=50$, $P_c/P_p =1.5$, and depict the results in Fig.~\ref{MIMO_long}. It is clear that the proposed solution has a superior performance for this increased blocklength as well. We note that the blocklength employed in the proposed algorithm is $3$ times shorter than those used in \cite{decurninge2020tensor,fengler2020pilot}. Using a larger blocklength would improve the system performance at the cost of higher computational complexity. To show the effect of the parameter $n_p$ on the performance of the proposed scheme, we provide several examples for $n_c = 512, J=2,M=50$, $P_c/P_p =1.5$ in Fig.~\ref{figDiffJ}. It is observed that the performance of the decoder is highly sensitive to this parameter, especially, for larger values of $K_a$.\\
\indent In Fig. \ref{detectorLong}, the performance of the Neyman-Pearson detector is shown for two different values of $n_p$ and $M$. It is demonstrated that if we select the value of $P_{p}$ greater than $0.005$ and $0.02$, the probability of detection is around $1$. Since the values of $P_{p}$ selected for reaching the target PUPE in Figs.~\ref{MIMO_short} and \ref{MIMO_long} are greater than these values, we benefit from the excellent pilot detection performance in these scenarios. It is also evident from these results that the detection probabilities obtained by simulations match the analytical result in \eqref{pdetect}.\\
\indent Finally, we note that a new scheme called FASURA has been reported in \cite{Gkagkos2022Fasura} after the submission of this paper. In FASURA, each user transmits a large blocklength signal containing a non-orthogonal pilot and a randomly spread polar code. FASURA offers improved performance for large blocklengths while our scheme remains superior for short blocklengths.
	\vspace{-.500 mm}
\section{Conclusions}
	\vspace{-.800 mm}
We propose an unsourced MAC scheme for block fading channels using a massive MIMO structure. The proposed scheme uses multiple stages of orthogonal pilots for pilot detection and channel estimation. The use of small-length orthogonal multi-stage pilots makes the system implementable for short blocklength scenarios. The results demonstrate that the proposed approach is superior to the existing alternatives developed in the recent literature.
 \vspace{-.5mm}
\begin{appendices}
 \vspace{-1mm}
 \section{Performance of the NP Hypothesis Testing}
 \vspace{-.5mm}
\label{appendixC}
\indent The likelihood ratio for \eqref{hypothesis} is given by
\begin{align}
    L(\mathbf{u}_{ji}) = \dfrac{\mathbb{P}\left(\mathbf{u}_{ji}|\mathcal{H}_1\right)}{\mathbb{P}\left(\mathbf{u}_{ji}|\mathcal{H}_0\right)}=\dfrac{1}{\sigma^M_1}e^{\mathbf{u}_{ji}^H\mathbf{u}_{ji}/\sigma_0^2},
\end{align}
where $\sigma_0^2 =  \dfrac{2(1+m_{ij}n_{p}P_{p})}{m_{ij}n_{p}P_{p}}$. Thus, the Neyman-Pearson test for detection of $\mathbf{\bar{b}}_{i}$ is obtained by  
\begin{align}
\delta_{NP}(\mathbf{u}_{ji}) &= \left\{\begin{matrix}
1&L(\mathbf{u}_{ji}) \geq \tau_0\\ 
0&L(\mathbf{u}_{ji}) < \tau_0
\end{matrix}\right.= \left\{\begin{matrix}
1&\mathbf{u}_{ji}^H\mathbf{u}_{ji} \geq \tau_0^\prime\\ 
0&\mathbf{u}_{ji}^H\mathbf{u}_{ji} <\tau_0^\prime
\end{matrix}\right.,\label{decisionRule}
\end{align}
where $\tau_0^\prime = \sigma_0^2\mathrm{Ln}(\tau_0\sigma^M_1)$. The false-alarm probability of the above decision rule is calculated as
\begin{align}
    P_{F}(\delta_{NP}) = \mathbb{P}\left(\mathbf{u}_{ji}^H\mathbf{u}_{ji} \geq \tau_0^\prime|\mathcal{H}_0 \right)
\numeq{a}  1-\Gamma_{2M}(2 \tau_0^\prime), \label{pFA}
\end{align}
where ($a$) follows the fact that $\mathbf{u}_{ji}^H\mathbf{u}_{ji}|\mathcal{H}_0\sim 0.5\chi^2_{2M}$ and $\mathbf{u}_{ji}^H\mathbf{u}_{ji}|\mathcal{H}_1\sim 0.5\sigma_1^2\chi^2_{2M}$, and $\chi^2_{k}$ denotes the chi-squared distribution with $k$ degrees of
freedom. To find the threshold for a $\gamma-$level Neyman-Pearson test, the probability of the false-alarm in \eqref{pFA} must satisfy $P_{F}(\delta_{NP})\leq \gamma$. Therefore, the threshold in \eqref{decisionRule} is obtained as
\begin{align}
    \tau_0^\prime= \dfrac{1}{2}\Gamma^{-1}_{2M}(1-\gamma).\label{threshold}
\end{align}
\indent The probability of detection in the absence of collision ($m_{ij}=1$) is then obtained as
\begin{align}
\nonumber
P_{D}(\delta_{NP}) =& \mathbb{P}\left(\mathbf{u}_{ji}^H\mathbf{u}_{ji} \geq \tau_0^\prime|\mathcal{H}_1 \right)\\\nonumber
=& 1-\Gamma_{2M}\left(\dfrac{2\tau_0^\prime}{1+n_{p}P_{p}} \right) \\
\label{pdetect}
=& 1-\Gamma_{2M}\left(\dfrac{\Gamma_{2M}^{-1}(1-\gamma)}{1+n_{p}P_{p}} \right) .
\end{align}

%\begin{align}
%\nonumber
%P_{D}(\delta_{NP}) =& \mathbb{P}\left(\mathbf{u}_{ji}^H\mathbf{u}_{ji} \geq \tau_0^\prime|\mathcal{H}_1 \right)\\\nonumber
%=& 1-\Gamma_{2M}\left(\dfrac{2\tau_0^\prime}{1+n_{p}P_{p}} \right)
%\end{align}
%\label{pdetect}
%\begin{align}
%= 1-\Gamma_{2M}\left(\dfrac{\Gamma_{2M}^{-1}(1-\gamma)}{1+n_{p}P_{p}} \right) .
%\end{align}
\end{appendices}

\end{document}